# Electron Correlation driven Metal-Insulator transition in Strained and Disordered VO$_2$ films

*A. D'Elia[1,*], C. Grazioli[1], A. Cossaro[1,2], B.W. Li[3], C.W. Zou[3], S.J. Rezvani[4,1], N. Pinto[5],*

*A. Marcelli[4,6,7], M. Coreno[7]*

1. IOM-CNR, Laboratorio TASC, Basovizza SS-14, km 163.5, 34149 Trieste, Italy;
2. Department of Chemical and Pharmaceutical Sciences, University of Trieste, via L. Giorgieri 1, Trieste, Italy
3. National Synchrotron Radiation Laboratory, University of Science and Technology of China, Hefei 230029, P. R. China
4. Istituto Nazionale di Fisica Nucleare, Laboratori Nazionali di Frascati, 00044 Frascati, Italy;
5. Scuola di scienze e tecnologie, Sezione di Fisica, Università di Camerino, Via Madonna delle Carceri 9, 62032 Camerino, Italy.
6. Rome International Centre for Material Science Superstripes, RICMASS, Via dei Sabelli 119A, 00185 Rome, Italy
7. ISM-CNR, Istituto Struttura della Materia, LD2 Unit, Basovizza Area Science Park, 34149 Trieste, Italy

*Corresponding author: delia@iom.cnr.it

## Abstract

The Metal-Insulator transition (MIT) in VO$_2$ is characterized by the complex interplay among lattice, electronic and orbital degrees of freedom. In this contribution we investigated the strain-modulation of the orbital hierarchy and the influence over macroscopic properties of the metallic phase of VO$_2$ such as Fermi Level (FL) population and metallicity, i.e., the material ability to screen an electric field, by means of temperature-dependent X-ray Absorption Near Edge Structure (XANES) and Resonant Photoemission spectroscopy (ResPES). We demonstrate that the MIT in strained VO$_2$ is of the Filling Control type, hence it is generated by electron correlation effects. In addition, we show that the MIT in Nanostructured (NS) disordered VO$_2$, where the structural phase transition is quenched, is driven by electron correlation. Therefore a fine tuning of the correlation could lead to a precise control and tuning of the transition features.

## Introduction

The ability to manipulate the charge carrier flow within a solid state material is at the basis of the modern electronics. There have been a vast investigations on such manipulations via the material dimension [1,2], surface-interface customization [3,4] and the structural strain application [5–7], however all compounds that intrinsically exhibit alternating itinerant-to-localized behaviour of the electrons result extremely appealing.

Among the different systems offering this possibility, such as perovskites, transition metal dichalcogenides and transition metal oxides [8–10] the $VO_2$ is one of the most studied.

$VO_2$ is a $3d^1$ electron system, which undergoes a reversible Metal-Insulator Transition (MIT) coupled to a structural phase transition, going from a low temperature monoclinic insulator to a high temperature tetragonal (rutile) metal [11,12]. The near room temperature (~340 K in bulk) phase transition, is of paramount interest for a variety of applications, ranging from energy saving and smart windows [13] to the emerging field of Mottronics [14,15].

The physical mechanism behind the insulating gap opening has been long debated. In literature, different models have been proposed to explain the $VO_2$ MIT: a structurally driven phase transition triggered by a Peierls instability [12], a pure electronic Mott-Hubbard phase transition [16] or a combination of both [17,18].

This long standing debate has been fuelled by the complex interplay between, lattice, orbital and electronic degrees of freedom, which modify the MIT features like transition temperature and width [5,7,19–28]. Among the different methods developed to control the MIT, the application of the epitaxial strain to $VO_2$ thin films emerged as one of the most effective ways to enhance the electron-electron interaction and, therefore, to customize the transition process [21,29–32]. However, complete disentanglement of electronic and structural degrees of freedom is not achieved in strained samples where the local atomic configurations deeply affect the orbital ordering of the metallic phase at the Fermi Level (FL) [5,21,27].

We used X-ray Absorption Near Edge Structure (XANES) and Resonant Photoemission spectroscopy (ResPES), to clarify the role of the local arrangements around vanadium atoms and the electron correlation over the orbital hierarchy, metallicity and Fermi Level population in the conductive phase of vanadium dioxide. Our research shows that the MIT in strained $VO_2$ is of the Filling Control (FC) type and as a consequence is mainly driven by electron correlation effects [33]. This is confirmed by studying the extreme case of a Nanostructured (NS) disordered sample where the structural phase transition is quenched by disorder, i.e. electronic and structural degrees of freedom are disentangled. In this limiting case, concurrent spectroscopic-electronic transport measurements point out the pure electronic character of the transition, suggesting that the electron interaction is the key parameter to tune MIT features.

## Experimental

Films of $VO_2$ having a thickness of 8, 16 and 32 nm were deposited on a clean substrate of $TiO_2$ (001) by the RF-plasma assisted Molecular Beam Epitaxy instrument working with a base pressure better than $4x10^{-9}$ mbar. These film were grown at the constant growth rate of 0.1 Å/s and their thickness was controlled by monitoring the deposition time. The substrate has been kept at the temperature of 550° C during the deposition process. The interfacial cross-section has been investigated with the high-resolution scanning transmission electron microscope (STEM). High angle annular dark field (HAADF) STEM images were taken on a JEM ARM200F with a probe aberration corrector, while the diffraction pattern was acquired on a JEM 2100 TEM. Information and details of the epitaxial films preparation are reported elsewhere [21,34].

The NS VO$_2$ sample has been synthetized using a Pulsed Micro-plasma Cluster Source available at TASC laboratories [35]. The PMCS is a pulsed-cluster source driven by a high-power pulsed electric discharge. In the present experiment, the PMCS was operated with a vanadium cathode (6 mm diam. rod, purity 99.9 %, EvoChem GmbH) generating a supersonic beam of vanadium oxide clusters. To obtain homogeneously oxidized nanoparticles, we used Ar (high purity Ar: 99.9995%, SIAD) as the carrier gas, mixed with a controlled amount of oxygen (O$_2$/Ar=0.3% mol) as described in [36]. The base pressure of the experimental chamber was $1*10^{-7}$ mbar, during the deposition process. The working parameters of the PMCS have been adjusted in order to maximize the deposition rate (delay between gas injection and discharge firing = 0.61 ms; discharge operating voltage 0.9 kV; discharge duration 60 μs; pulsed-valve aperture driving signal duration time 210 μs; pulse repetition rate 3 Hz; Ar-O$_2$ pressure feeding the Parker valve 60 bar).

The XANES and ResPes experiments have been performed at the ANCHOR end-station of the ALOISA beamline [37] at Elettra synchrotron radiation facility. Electrons were collected at normal emission by a PSP Vacuum 120 mm electron analyser with 2D delay line detector. The photon beam was linearly polarized in the scattering plane and impinging the sample at the magic angle (35° measured respect to the sample surface). Measurements were performed at constant pass energy ($E_p$=20 eV).

The XANES spectra have been acquired in Auger yield (O $KL_{23}L_{23}$ ~ 507 eV) in order to minimize the V L edges contribution [38,39].

Transport properties have been carried out at dark in a small furnace operating in vacuum (P < $10^{-5}$ mbar). The sample temperature has been measured and controlled within 0.1 K by a thermocouple (type N) connected to the temperature controller Eurotherm mod. 3216. The film temperature has been changed slowly, with a rate of ~ 0.5 ÷ 1 K/min, in an uninterrupted cycle from room temperature (RT) up to about 400 K and then back to RT. The two co-planar contacts geometry has been chosen, considered the high RT resistance exhibited by all films, fixing two thin Cu wires onto the film surface by using silver paint. The film resistance has been measured by an electrometer Keithley 6517B operated in the V/I mode, applying a constant bias (0.5 V) and measuring the current flowing through the contacts. Each R(T) value has been measured after stabilization of the temperature better than 0.1 K. For every T, at least 25 points of R(T) values have been averaged to calculate the corresponding ρ(T) mean value. For each film the measurement has been repeated several times under the same nominal conditions, but the applied bias [40,41]. Current-voltage (I-V) characteristics have been detected at several fixed temperatures, by using a source-meter Keysight mod. B2912A.

## Results and Discussion

To recognize the multi-orbital character of the MIT it is necessary to understand how the local arrangement of the oxygen atoms around the vanadium atom determines the band structure of VO$_2$. In the metallic phase of bulk vanadium dioxide, ligand atoms arrange according to a slightly distorted octahedral symmetry around the metal site with two nonequivalent V-O bond distances, namely equatorial and apical [5,21,31]. The octahedral crystal field splits the degenerate 3d manifold into 3 $t_{2g}$ and 2 $e^\sigma_g$ levels. In addition, the small orthorhombic distortion separates the 3 $t_{2g}$ levels in one singly degenerate $a_{1g}$ and two $e^\pi_g$ levels. The $e^\pi_g$ and $e^\sigma_g$ orbitals hybridize with the O 2p, forming bonds of σ and π symmetry [12]. Their unoccupied levels are identified as π*($e^π_g$ character) and σ* ($e^σ_g$ character). The $a_{1g}$ orbital, populated by unpaired *3d* electrons, it is called $d_∥$ [12]. In the conductive state, π* and $d_∥$ bands are slightly degenerate at the Fermi Level (FL).

In the insulating phase, the pairing of vanadium atoms increases the superposition of V *3d* wave functions splitting the $d_∥$ creating an antibonding empty $d_∥^*$ band. Within the ligand octahedron, the V-V dimerization has the concomitant effect of moving off-centre the metal atom and increasing the superposition between $e^π_g$ and oxygen wave functions. As a consequence the π* band is upshifted in energy opening a gap. The antibonding $d_∥^*$ is strictly related to the presence of unidimensional V-V dimer chains in the monoclinic insulating phase [42].

Changing the atomic distances in a controlled fashion, e.g. through the application of an epitaxial strain, affect both wave functions overlap and electron correlation and, as a consequence, the electronic structure of $VO_2$. The lattice mismatch of $VO_2$ grown on $TiO_2$ (001) thin films results in an increase of the apical V-O distance, which reduces the superposition between oxygen and vanadium orbitals and therefore the 3d-2p hybridization [21,31]. The π* orbital, which points toward other vanadium atoms and it is directed between the oxygen corners of the octahedron, is the most affected [5]. As a consequence, the decrease in V-O hybridization reduces the bonding-antibonding energy separation, hence the π* orbital is shifted to lower energy. At variance, the $d_∥^*$ shifts upwards since the strain reduces the inter-pair distance increasing the orbitals superposition within the unidimensional V-V chains. In synthesis, in the metallic phase, the strain reverses the π*-$d_∥^*$ hierarchy modifying the orbital population at the FL. In the insulating phase where the V-O hybridization is stronger, the π*-$d_∥^*$ splitting increases, with $d_∥^*$ being at higher energy. Increasing the wave functions overlap also increases the electron correlation, which has been demonstrated to have a major effects on the MIT in $VO_2/TiO_2$ (001) thin films respect to the strain-induced structural rearrangement [5].

For the NS disordered sample the description of the electronic structure is complicated by the presence of multiple distorted phases coexisting within the same film. As demonstrated by a previous investigation, the $VO_2$ NS sample synthetized by a PMCS is characterized by a local disorder [43]. The disordered nature of this NS $VO_2$ film point out that the role of the Peierls pairing mechanism that requires an ordered crystalline structure can be neglected, giving us the unique opportunity to study a $VO_2$ sample in which structural and electronic contributions to the phase transition are almost disentangled.

To confirm the quality of the samples and the resistive switching behaviour of $VO_2$ transport measurements were carried out on samples of 8, 16 and 32 nm thickness as described in the experimental section in the cooling-heating uninterrupted cycle. The resistivity behaviour of these samples as a function of temperature is shown in Figure 1a. The resistivity curves show the typical hysteresis observed in $VO_2$, while the transition temperature shifts with the film thickness from ~315 K in the 8 nm film to ~ 345 K in the 32 nm film in agreement with the relative $VO_2$ bulk transition temperature [11,44] . The decrease of the temperature can be directly correlated to the strain induced variation of the V-V/V-O distances and hence to the variation of the p-d hybridization and electron-electron interaction as will be discussed in details later on. The resistivity temperature coefficient derived as -1/ρ(dρ/dT) with a numeric derivative of the resistivity as function of the temperature, returns a transition width of the order of 15 K (see figure 1b), indicating a high degree of crystallinity in samples [45].

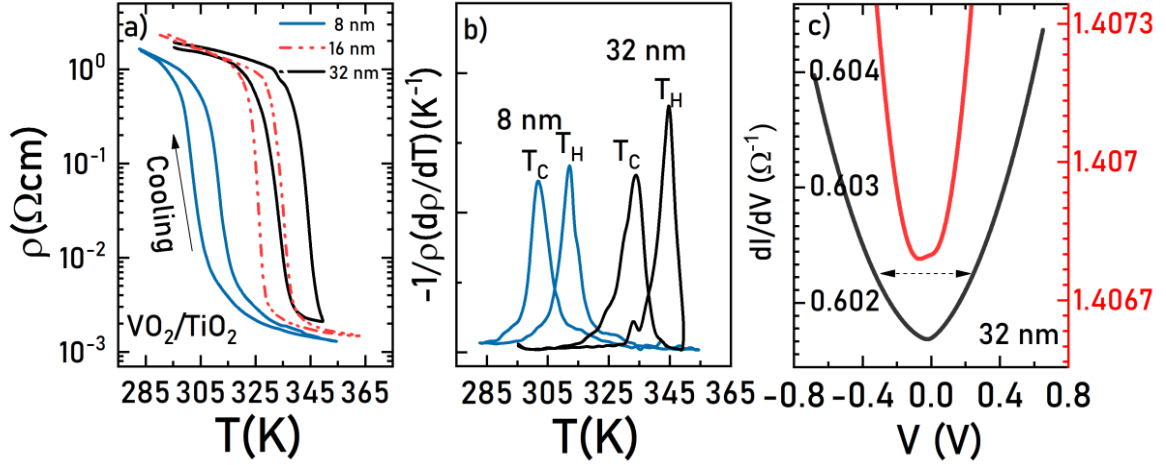

*Figure 1: The transport properties of the VO₂ thin films: a) the resistivity of 8, 16 and 32 nm thick film as a function of temperature in the heating-cooling cycle; b) the temperature coefficient of the resistivity indicating the transition width; c) the differential conductivity of the 32 nm sample at 300 (black) and 365 (red).*

The differential conductivity, σ, of the 32 nm films calculated from the numeric derivative of the I-V measurements at 300 and 365 K is shown in figure 1c. The calculated quantity is directly correlated to the transition available density of states (DOS) in which the voltage widths around and close to zero is proportional to the insulating gap of the band structure. Thus, the voltage width reduction in the dσ curve of the 32 nm thick film, increasing the temperature from 300 K to 365 K, points to a transition to a metallic phase with an increase of the empty states. The decrease of the dynamic resistance width starts as soon as the resistance value in the split region of the curve of the resistivity vs. temperature decreases, confirming the hypothesised shift in the DOS and the change of the population of π*, $d^*_{\parallel}$ bands at the FL as a function of the temperature.

In order to correlate the MIT features with the orbital ordering across the phase transition, a series of temperature dependent XANES measurements have been performed. In vanadium oxides, V *3d* and O *2p* electrons strongly hybridize [24] hence measuring the O K edge XANES [38,44,46] it is possible to probe the π*, $d^*_{\parallel}$ and σ* bands and, in great detail the empty bands with $t_{2g}$ and $e_g$ character.

Figure 2 reports the XANES spectra of the VO₂ films. The 8, 16 and 32 nm films show intense and well-separated π* and σ* resonances at about 529.8 and 532.2 eV, respectively indicating an intense crystal field contribution that indicates also the local order of these samples [38,46]. At variance, the $d^*_{\parallel}$ band sets between 530.5 and 531 eV is not well resolved [29,44,46]. Across the MIT a slight increase of the intensity in the π* energy region is observed (see Figure 2 left panel) concurrent with a depletion of the valley between π* and σ*, which going from the high to low temperature phases is compatible with the collapse of the $d^*_{\parallel}$ band at the FL.

On the other hand the nanostructured film characterized by a disordered nature does not well separate π* and σ* resonances [43,46,47]. This interpretation is also supported by the shift of the σ* band toward a lower photon energy by ~1 eV respect to the crystalline samples. σ* relative position respect to π* is directly related to the 10Dq crystal field intensity [48] and therefore to the local structural order of the ligands arrangement. Nevertheless across the MIT, spectral changes are observed also in the $t_{2g}$* energy region.

For strained samples, decreasing the thickness we observed an increase in the deepness of the $d^*_{\parallel}$ valley across the MIT. This is compatible to a hierarchical modulations of the π* and $d^*_{\parallel}$ bands induced by the epitaxial growth [5,21].

To investigate the strain induced orbital hierarchy inversion, we investigated the differences among the XANES spectra between the insulating and the metallic phase.

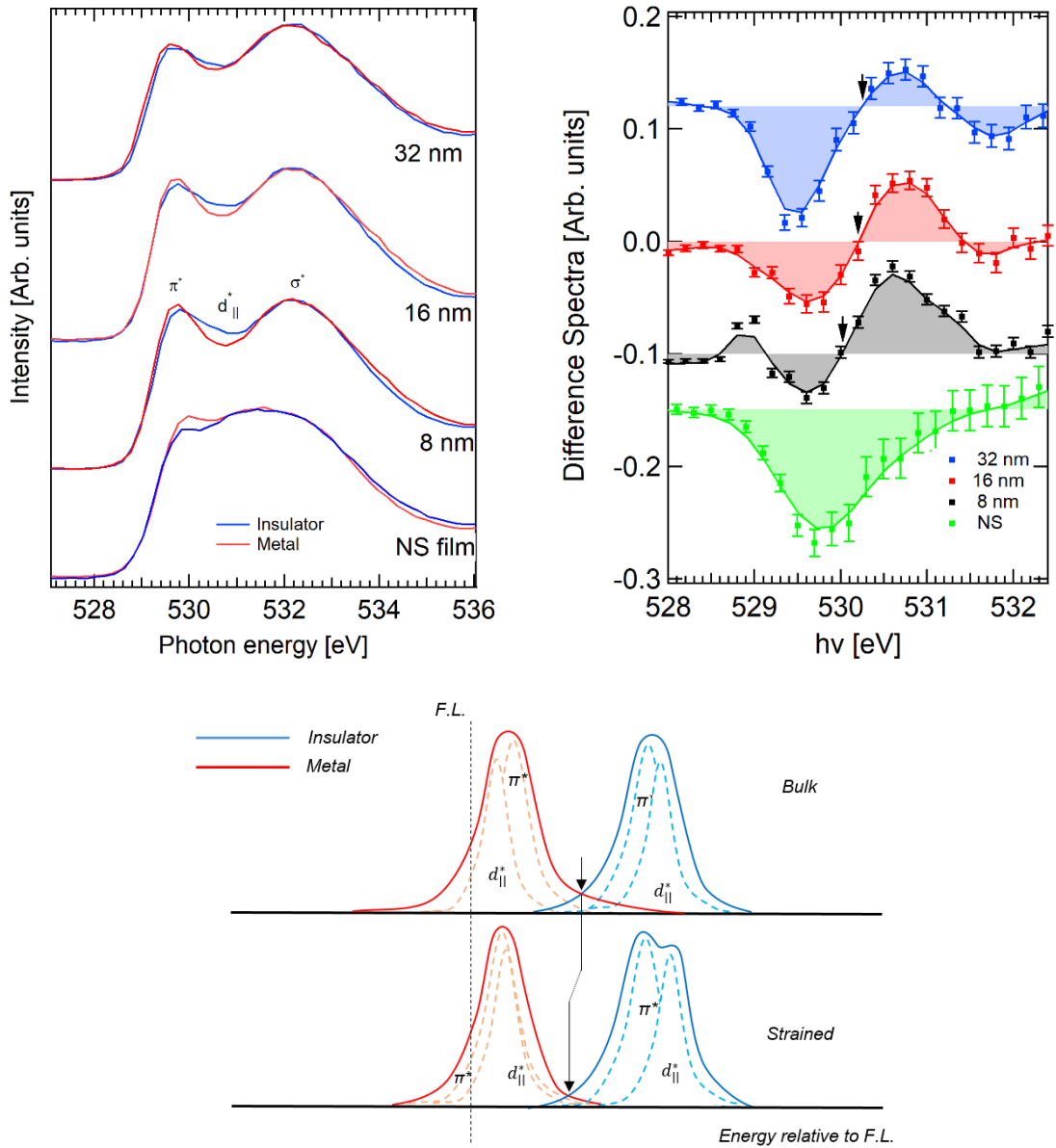

*Figure 2: Left panel: comparison of the O K edge XANES (527-536 eV) of VO$_2$ crystalline film of 32, 16 and 8 nm and the NS disordered film for metallic (red, 90° C) and insulating (blue, 30° C) phases. XANES spectra are normalized respect to the maximum intensity of the σ* feature at 532.2 eV. Right panel: comparison of the difference spectra of the O K edge XANES in the range hv: 528-532 eV. The dots represent experimental points while the continuous line (a guide for the eye) is the smoothed curve of the experimental points (binomial algorithm). From top to bottom: films of 32, 16 and 8 nm and the NS film. The black vertical arrows indicate the turning point of the difference from negative to positive contributions. Spectra are vertically shifted for sake of clarity. Bottom panel: the schematic empty band model for bulk and strained VO$_2$ in the insulator (blue) and metal (red) phases. The intersection point (marked by arrows) between the insulator and metal empty DOS shifts toward lower energy for the strained sample, because of the concurrent π*, $d^*_{\parallel}$ separation in the insulating phase and π*, $d^*_{\parallel}$ inversion in the metallic phase.*

The differences among XANES spectra are shown in the right panel of Figure 2 and have been calculated as [31]:

$$I_{ins} - I_{met} \propto uDOS_{ins} - uDOS_{met} \qquad (1)$$

where $I_{ins}$ and $I_{met}$ are the intensity of the XANES spectra of the insulating and the metallic phases and $uDOS_{ins}$ and $uDOS_{met}$ are the empty DOS, respectively. Using the Eq. 1 a negative contribution is expected in the spectral region where the metallic empty DOS is more intense. A turning point (TP) from negative to positive contribution is also expected where the metallic and the insulating empty DOS coincide. In our samples as the strain increases the positive contribution in the *hv* range 530-531 eV grows with an asymmetry toward high photon energy. The negative contribution in the 528-529.7 eV decreases and the TP shifts toward lower photon energy. The latter is a fingerprint of the shrinking of the metal empty DOS with a simultaneous widening of the insulating empty DOS. They are expected to cross at a lower photon energy, in agreement with the qualitative band description of Figure 2 and the strain dynamics of π*-$d_{\parallel}^{*}$ bands in insulating and metallic phases. The growing asymmetry in the positive contribution at 530-531 eV, is also in agreement with the strain-induced increase of the π*and $d_{\parallel}^{*}$ separation in the insulating phase. Furthermore, increasing the film strain, the negative contribution decreases which is a result of a highly occupied π* band in the strained films.

Looking at the XANES spectra of the metallic phase, decreasing the thickness (i.e. increase of the strain) is evident the increase of the Filling Ratio F=(π*+$d_{\parallel}^{*}$)/ σ*. Indeed, both π*and $d_{\parallel}^{*}$ are partially degenerate crossing the FL in the metallic phase. Considering that the σ* contribution to the XANES spectra it is almost independent from the strain due to the stability of the σ bond, an increase of F, reveals an increase of the empty states with π* and $d_{\parallel}^{*}$ character. This implies a decrease of the number of occupied states and hence a decrees of free carriers in the most strained sample.

This demonstrate that the inversion in the π*-$d_{\parallel}^{*}$ orbital hierarchy is concurrent with a depletion of the FL population in the metallic phase of $VO_2$.

The FL filling dynamic as a function of sample thickness clearly points out that the MIT is of the FC type, suggesting the purely electronic character of a transition dominated by the short-ranged electron correlation rather than the electron–lattice coupling [33].

Regarding the spectral difference of the NS film, in Figure 2 it is evident the negative contribution whose minimum is aligned with the strained samples. Going from the low temperature to the high-temperature phase, an increase of the intensity in the energy region of the $t_{2g}$* is observed, pointing out that the main spectral changes across the MIT affect the empty orbitals with $t_{2g}$ character.

To study the effect of local order over the electronic structure of $VO_2$ a set of ResPES measurements have been also performed exploring the photon energy through the V $L_3$ edge. Spectra of the 8, 16, 32 nm samples and of the disordered NS sample in the -2; 3 eV Binding Energy (BE) region collected at the maximum of the V $L_3$ resonance are compared in Figure 3.

For the entire set of samples, spectra of the metallic state are characterized by the population at the FL and are comparable with previously published data for $VO_2$ thin films [49]. The maximum of the *3d* signal enhancement is observed at 518.4±0.1 eV, in agreement with the maximum of the XANES V L edge spectrum [31], confirming once more the *3d* nature of the 1-2 eV peak.

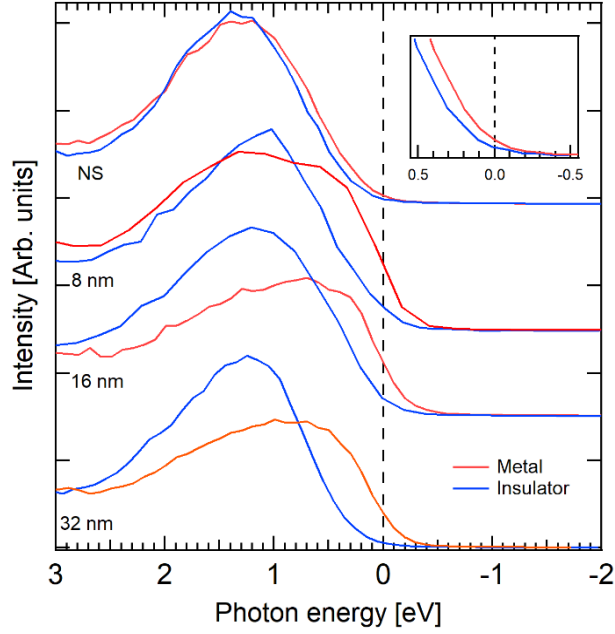

*Figure 3: VB spectra of the unpaired 3d electrons band in the BE range (-2; 3) eV in the insulating (blue) and metallic (red) phase of VO$_2$. From top to bottom: the NS disordered sample, samples 8, 16 and 32 nm thick. The inset shows the increase of the population at the FL for the NS sample.*

We used ResPES as a probe of the metallicity (i.e. the ability to screen an electric field) in our set of samples [50–52]. The maximum intensity (normalized to the photon flux) of resonant photoemission feature can be linked to the screening parameter by the Eq. 2 [50–52]

$$\frac{I}{I_0} \alpha \frac{1}{L_s} \qquad (2)$$

where *I* is the ResPES peak intensity, $I_0$ the incident photon flux and $L_s$ the screening parameter of the material (the inverse of the Thomas-Fermi screening length [52]). In this framework, in an insulating material, (small $L_s$), the core-hole generated by the ResPES process is poorly screened allowing the resonantly excited state to decay in order to interfere with the direct photoemission. On the other hand, the delocalized electrons inside a metal (high $L_s$) quickly screen the photo-hole partially quenching the valence-to-core decay of the excited state, reducing the enhancement of the resonant signal.

When the excitation energy *hv* is set at the maximum of the vanadium $L_3$ threshold, the insulating *3d* peak is always more intense respect to the metallic counterpart as observable in Figure 3. Applying Eq. 2 to our dataset, we may calculate the ratio between the insulating and the metal phase ResPES intensity obtaining $L_m/L_i$ (the ratio of the screening parameters of the metallic and insulating phases) as a function of the sample thickness (Figure 4). The $L_m/L_i$ values extracted are: 1.93, 1.73 and 1.44 for the 32,16 and 8 nm samples respectively.

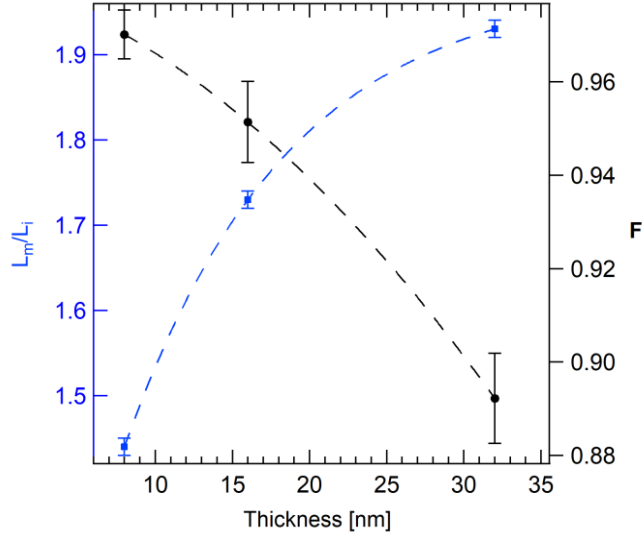

Figure 4: Ratio of the ResPES intensity in the two phase: $L_m/L_i$ (left axis) and the Filling ratio $F=(\pi^*+d^*_{||})/\sigma^*$ (right axis) as a function of the sample thickness.

$L_m/L_i$ can be estimated considering that the screening parameter is the inverse of the Debye screening length:

$$L_s = \frac{1}{\lambda_D} = \sqrt{\frac{e^2 n}{\varepsilon k_b T}} \quad (3)$$

Where $\lambda_D$ is the Debye screening length, $e$ is the carrier charge, $n$ is the carrier density, $\varepsilon = \varepsilon_0 \varepsilon_r$ is the average dielectric constant, $k_b$ the Boltzmann constant and T the temperature. Therefore the ratio of the screening parameter of the metallic to insulator phase is:

$$\frac{L_m}{L_i} = \sqrt{\frac{n_m \varepsilon_i T_i}{n_i \varepsilon_m T_m}} \quad (4)$$

Where the subscript *m* indicate the quantity relative to the metallic phase and *I* those relative to the insulating phase. Using the values of the dielectric constant reported by Zheng and coworkers [53] ($\varepsilon_i=40$ and $\varepsilon_m=10^4$) and the carrier concentration $n_i=1.9*10^{19} cm^{-3}$ $n_m=1.9*10^{23} cm^{-3}$ [54], with temperature $T_i=303.15$ K and $T_m=363.15$ K we obtain $L_m/L_i \approx 5.7$ for bulk $VO_2$.

This value is not consistent with the measured $L_m/L_i =1.93$ obtained for the 32 nm thick, bulk-like samples. This is mainly due because the Debye model cannot properly evaluate $L_m$. In fact, it is necessary to consider that the conductive phase of $VO_2$ is a "poor" metal in which electron correlation effects drive the transport properties. Indeed, metallic vanadium dioxide violates the Ioffe-Regel-Mott criterion [55], or in other words, the Boltzmann statistic cannot be used to describe $VO_2$ properties and therefore the Eq. 4 is a poor approximation of the metallic screening. Nevertheless, the Debye model allows to link the screening parameters ratio $L_m/L_i$ to the carrier density of the metallic phase $n_m$, allowing us to interpret its thickness dependent variation as mainly due to the concurrent variation of the carrier density in the metallic phase.

In Fig. 4 we show the trend of $L_m/L_i$ (blue dots) as a function of the sample thickness vs. F, the Filling ratio. $L_m/L_i$ and F have opposite behaviors thus confirming the previous interpretation. In the metallic phase the strain affects the population at the FL, which is depleted respect to the bulk, i.e., F is higher in the most strained sample. Concurrently, due to the lack of itinerant electrons the screening parameter decreases in the strained samples. In other words, decreasing FL population, increases the spectral weight of the empty DOS (increase of F), this transfer of spectral weight is an additional fingerprint of the FC nature of the MIT in $VO_2$ [33].

This result combined with the XANES analysis, unambiguously links the strain-induced orbital hierarchy inversion in the metallic phase of $VO_2$, with the variation of population at the FL and the metallicity, determining the FC nature of the MIT, i.e., driven by the electron correlation.

In the limiting case of the NS sample the appearance of a small, but finite DOS at the FL in the high-temperature spectra of the VB (inset of Figure 3) points out the increase of the metallicity from 303.15 to 363.15 K, in agreement also with the value of the ratio $L_{metal}/L_{insulator}$ >1 (see Supplementary materials). In order to rule out a purely thermal mechanism for the modulation of the population at the FL, resistance measurements have been performed. As shown in Figure 5 the temperature dependence of the resistance does not obey the Arrhenius law [56] and, since the Peierls mechanism is quenched by disorder, the electron-electron interaction is the only possible mechanism responsible of the change of the population at the FL.

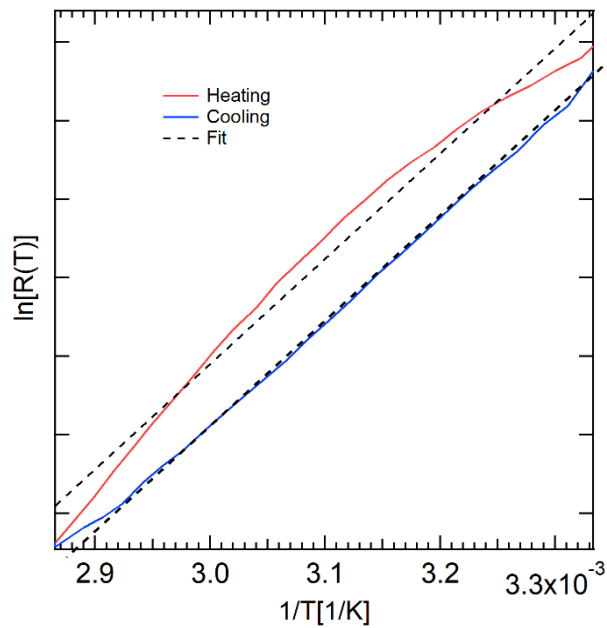

Figure 5: The Arrhenius plot of the resistance of the NS disordered sample in the temperature range 300-350 K. Experimental data deviate from the theoretical linear dependence.

A straightforward proof of the purely electronic character of MIT in NS $VO_2$ film is achieved through Constant Initial State (CIS) spectroscopy. CIS is an advanced spectroscopy technique, in which the modulation of the intensity of a photoemission feature is recorded as a function of the photon energy [38,39,57]. Respect to XANES, CIS spectroscopy is highly selective. In a CIS spectrum the photon energy and the kinetic energy of the electron are simultaneously varied to fulfil the condition with the ionization energy ($E_I$) as a constant.

$$E_I = h\nu - E_{KE} \tag{5}$$

Since $E_I$ (or BE) is fixed, CIS measures the cross section ResPES lineshape mapping only the empty DOS region, which allows a resonant decay channel in a final state with a hole at the chosen $E_I$.

Across the phase transition of $VO_2$ the major spectral changes happen around FL.

As observable in Figure 3 (and supplementary materials) in the metallic phase, the valence band line-shape in the 0-3 eV range, is determined by the contribution of two features centered at about ~1.5 and ~0.5 eV. The presence of two contributions near the Fermi level is the result of the strong electron-electron interaction (U) in Mott-Hubbard metals [58]. For historical reasons, the low BE feature has been called coherent (or quasi-particle) peak since it is obtained by band structure calculations in the framework of the independent particle approximation, neglecting the many-body effects, while the high BE feature is named incoherent peak [58].

The coherent peak is generated by delocalized 3d band states. The quasi-particle and the incoherent peak can be assigned to two different final states in which two screening mechanisms operate. After the photoemission of a *3d* electron, from the initial state with a formal $3d^1$ configuration the system goes in a $3d^0$ final state. According to Mossanek and Abbate, the $3d^0$ configuration is poorly screened by the surrounding atoms while features near the Fermi Level are well-screened configurations [59]. In the $VO_2$ metallic phase, the most favorable final configurations are $3d^1\underline{C}$ (coherent hole) and $3d^1\underline{L}$ (ligand hole) [59]. The $\underline{C}$ state accounts for a non-local screening channel of the photo-hole provided by the surrounding vanadium atoms. In the $\underline{L}$ configuration, the nearest oxygen atoms provide the charge to screen the missing electron. According to theoretical calculations, we assigned to the $3d^1\underline{L}$ configuration the insulating phase peak around 1.2 eV of BE while the 0.5 eV feature has $3d^1\underline{C}$ nature [49,59].

Considering this, we studied the CIS spectra of the V *3d* $\underline{C}$ and $\underline{L}$ features which are reported in Figure 6. The spectra have been acquired spanning the photon energy across the V $L_3$ resonance.

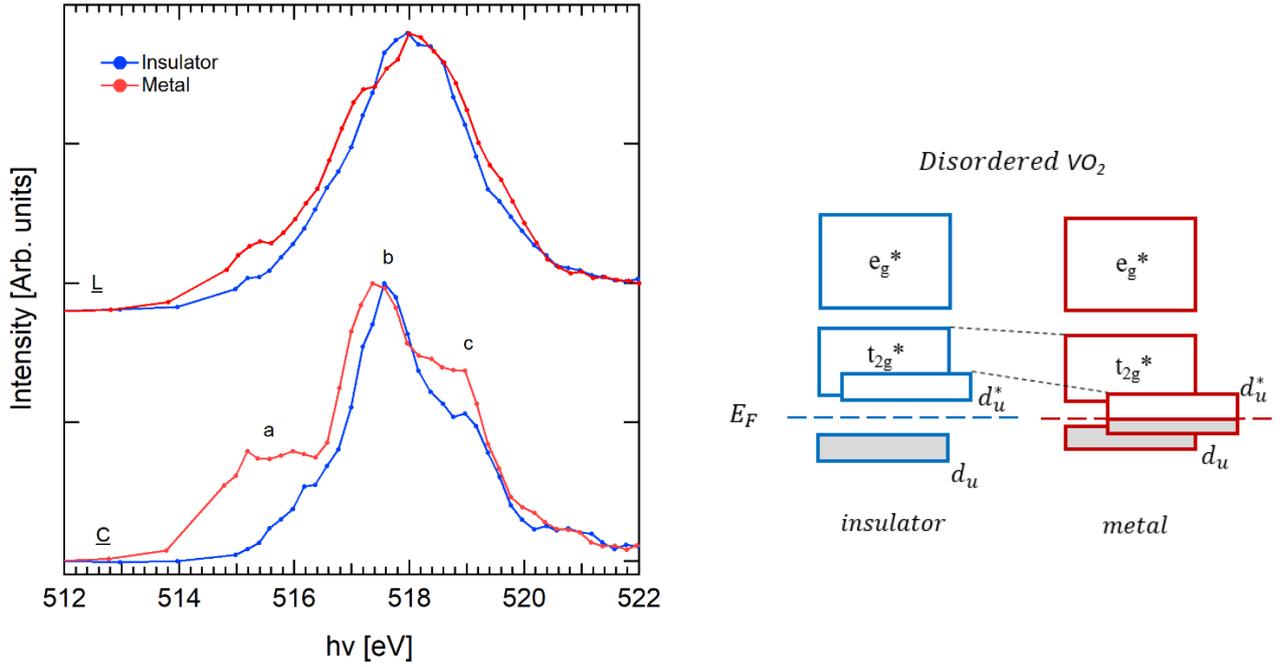

*Figure 6: Left panel, V 3d CIS spectra for L (BE=1.5 eV) and C feature (BE=0.4 eV) in the metallic (red) and insulating (blue) phase of the disordered NS sample. Spectra are normalized to the incident photon flux and to their maximum intensity. Right panel, schematic representation of the band evolution across the MIT in a disordered sample. The $d_u$ band, originated by unpaired 3d electrons, in the insulating phase splits in $d_u$ and $d_u^*$ bands by electron-electron repulsion. Across the phase transition $d_u^*$ collapses, crossing the Fermi energy and closing the band gap.*

The L CIS spectra strongly resemble the V $L_3$ XANES spectrum [43]. Minor spectral changes are observed across the phase transition at hv ~ 515 eV. This is not surprising since the L photoemission feature is connected with the local screening channel, i.e., is associated to well localized electrons. As a consequence since mostly *V 3d - O 2p* hybridized electrons are involved no major spectral changes across the MIT are expected.

On the other hand, the C photoemission feature is intrinsically linked to the conductive phase of $VO_2$. C CIS spectra select the excited states, i.e., the empty DOS region, for which the *V 3d* photo-hole is well screened by the delocalized metallic channel. In other words C CIS is sensitive to the empty DOS region, which contributes to the metallic phase of the $VO_2$. Three main features can be recognized: *a* (~ 515 eV), *b* (~517.6 eV) and *c* (~519 eV). The $e_g^*$ character can be assigned to the high-energy *c* feature while the $t_{2g}^*$ to the *b* feature. The *a* feature nature deserves a deeper discussion because while it is evident in the spectrum of the conductive phase, its presence can be only associated to the shoulder around 516 eV in the low temperature CIS spectrum.

The appearance of the *a* feature is accompanied by the increase of the *c* intensity. Since CIS spectra are normalized to the maximum of the *b* feature, any spectral change has to be read as a transfer of the spectral weight from *b* to *a*. As a consequence, the *a* feature can be interpreted as of one of the $t_{2g}$ empty orbitals not degenerate with the other two (*b* feature). The fact that *a* collapses toward low energy in the metallic phase while is partially degenerate with the $t_{2g}^*$ features in the insulating phase is the evidence of the *d*-band splitting across the MIT. Actually, the *a* feature in the CIS spectra can be associated to the $d_u^*$ band in the model in Figure 6 (right panel).

Since the Peierls distortion in the NS sample is quenched, in this case, the insulator to metal transition can only be driven by the electron-electron interaction and correlation [23,60] allowing us to determine the orbital evolution across the MIT in a disordered $VO_2$ system.

The behaviour in the disordered NS sample confirms the electronic nature of the MIT and strengthens our interpretation of a MIT driven by electron-electron interaction and of the FC type, substantially independent by the lattice degree of freedom.

An additional phenomenology of a FC transition is the occurrence of, charge ordering (CO) phases. CO can arise in a *$3d^n$* system when the filling number *n* is not integer [33]. Our study demonstrate that strain can be used to reduce the number of itinerant electrons, thus *n*, setting the necessary conditions for the emergence of CO in $VO_2$.

C. N. Singh and co worker recently reported CO phenomena in $VO_2$/$TiO_2$ thin films [61], strengthening our interpretation of the FC nature of the MIT.

## Conclusions

This study demonstrates that short-ranged electron interaction is the decisive parameter behind $VO_2$ MIT. The application of strain emerges as an invaluable tool to finely tune FL occupancy and metallicity in the metallic phase of $VO_2$.

Our spectroscopic analysis on strained and disordered samples, indicate that the MIT is of the FC type, therefore mostly independent from electron-lattice coupling. The identification of strain as a control parameter for band filling and occupation is of paramount importance since it can be used instead of solid state solutions or doping as the key to control the $VO_2$ MIT.

The role of strain over electron-electron interaction will enable the design of ultra-thin electronic devices with controllable electronic properties, such has band occupancy, based on the MIT, boosting the emerging field of Mottronics for efficient new generation devices. In addition, the discovery of strain as control parameter of FL occupancy and therefore of electron correlation, is likely to be influential also in transversal hot topic fields such as that of ultra-thin perovskite films.